\documentclass[twocolumn,prstab,floatfix]{revtex4}
\usepackage{graphicx}
\usepackage{amsmath}
\usepackage{latexsym}
\usepackage{amsfonts}
\usepackage{amssymb}

\begin{document}

\title{Synchro-Betatron Stop-Bands due to a Single Crab Cavity}
\author{Georg H.~Hoffstaetter} 
\affiliation{Department of Physics, Cornell University, Ithaca/NY}
\author{Alexander W.~Chao}
\affiliation{Stanford Linear Accelerator Center, Menlo Park/CA}

\begin{abstract}
We analyze the stop-band due to crab cavities for horizontal tunes
that are either close to integers or close to half integers. The
latter case is relevant for today's electron/positron colliders.  We
compare this stop-band to that created by dispersion in an
accelerating cavity and show that a single typical crab cavity creates
larger stop-bands than a typical dispersion at an accelerating cavity.

We furthermore analyze whether it is beneficial to place the crab
cavity at a position where the dispersion and its slope vanish. We
find that this choice is worth while if the horizontal tune is close
to a half integer, but not if it is close to an integer. Furthermore
we find that stop-bands can be avoided when the horizontal tune is
located at a favorable side of the integer or the half integer.

While we are here concerned with the installation of a single crab
cavity in a storage ring, we show that the stop-bands can be weakened,
although not eliminated, significantly when two crab cavities per ring
are chosen suitably.
\end{abstract}

\maketitle

\section{Introduction}
A crossing angle at the interaction region (IR) of colliding beam
accelerators introduces synchro-beta responses
\cite{piwinski77,piwinski85}. This is due to the fact that the
beam-beam focusing is different for particles at different positions
along the bunch, so that the horizontal beam-beam force is modulated
by synchrotron oscillations of the particles.  Crab cavities have been
envisioned to eliminate this source so coupling for linear colliders
\cite{palmer88} and circular colliders \cite{oide89}. While
construction work started in early work on B-factories
\cite{Padamsee:1991wv} it is now planned to use crab cavities for the
first time in KEK-B, in order to have bunches collide head on in IRs
with crossing angle \cite{Akai:2004np}. While this eliminates the
synchro-beta resonances due to the beam-beam kick if the hourglass
effect is negligible \cite{krishnagopal89}, it requires transverse
kicks in the crab cavities that depend on the longitudinal bunch
position and therefore, even in absence of beam-beam interaction, crab
cavities themselves introduce synchro-beta resonances. Furthermore it
has been suggested that crab cavities could be used in the damping
rings of a linear collider.

Consider a storage ring that has an accelerating cavity and a crab
cavity. The transport matrix $\underline T^{kc}$ from the accelerating
cavity to the crab cavity for the phase space coordinates $x$, $x'$,
$\tau$ and $\delta$ is given by the Twiss parameters at both places
and by the betatron phase advance $\phi_{kc}$ from the accelerating to
the crab cavity \cite{handbook},
\begin{eqnarray}
\underline T^{kc}
&=&
\left(\begin{array}{ccc}
\underline T^{kc}_x & \vec 0 & \vec d_{kc} \\
\vec t_{kc}^T       & 1      & T^{kc}_{56} \\
\vec 0^T       & 0      & 1
\end{array}\right)\ ,\nonumber\\
\underline T^{kc}_{x11}
&=&
\sqrt{\frac{\beta_k}{\beta_c}} [
\cos(\phi_{kc})+\alpha_c\sin(\phi_{kc}) ]\ ,\nonumber\\
\underline T^{kc}_{x12}
&=&
\sqrt{\beta_k\beta_c}\sin(\phi_{kc}) \ ,\nonumber\\
\underline T^{kc}_{x21}
&=&
\frac{(\alpha_c-\alpha_k)\cos(\phi_{kc}) -
(1+\alpha_k\alpha_c)\sin(\phi_{kc})}{\sqrt{\beta_k\beta_c}}\ ,\nonumber\\
\underline T^{kc}_{x22} &=&
\sqrt{\frac{\beta_c}{\beta_k}} [
\cos(\phi_{kc})-\alpha_k\sin(\phi_{kc}) ]\ .\nonumber
\end{eqnarray}
We use the coordinate $\tau=(t_0-t)\frac{E_0}{p_0}$ which is the
complex conjugate to $\delta=\frac{E-E_0}{E_0}$ where $t_0$, $E_0$ and
$p_0$ are the time of flight, the energy and the momentum of a
reference particle. For highly relativistic particles $\tau=z$, the
longitudinal position of the particle.  If the non-periodic dispersion
starts at the cavity with zero, it is $\vec d_{kc}=(D_{kc},D'_{kc})^T$
at the crab cavity. Since the transport matrix has to be symplectic,
$\vec t_{kc} = (T^{kc}_{51},T^{kc}_{52})^T$ is given by
\begin{equation}
\vec t_{kc} = -\underline T^{kc T}_x\underline J\vec d_{kc}\ .
\end{equation}

The transport matrix $\underline T^{cc}$ from after the cavity once
around the ring to just before the cavity is given by
\begin{equation}
\underline T^{cc}
=
\left(\begin{array}{ccc}
T^{cc}_x      & 0 & \vec d_c \\
\vec t_c^T    & 1 & -2\pi R\alpha -\vec t_c^T\vec\eta_c\\
\vec 0^T      & 0 & 1
\end{array}\right)\ ,
\end{equation}
where $\alpha$ is the momentum compaction factor and $2\pi R$ is the
storage ring's circumference.  The periodic dispersion
$\vec\eta_c=(\eta_c,\eta'_c)^T$ at the cavity determines
\begin{equation}
\vec d_c = (\underline 1-\underline T^{cc}_x)\vec\eta_c\ ,
\end{equation}
and symplecticity again requires
\begin{equation}
\vec t_{kc}
=
-\underline T^{cc T}_x\underline J\vec d_{kc}\ .
\end{equation}
This leads to the term $\vec t_c^T\vec\eta_c = -2\pi R\alpha h_c
\sin\mu$ with $\mu$ being the betatron phase advance per turn, and
\mbox{$h_c=\frac{1}{2\pi
R\alpha}(\gamma_c\eta_c^2+2\alpha_c\eta_c\eta_c'+\beta_c\eta_c^{'2})$}.

The transport matrix of the accelerating cavity is given by
\begin{equation}
\underline T_{\rm cav} = \left(\begin{array}{cccc}
1   & 0 & 0 & 0 \\
0   & 1 & 0 & 0 \\
0   & 0 & 1 & 0 \\
0   & 0 & \frac{2\sin^2(\frac{\mu_s}{2})}{\pi R\alpha} & 1
\end{array}\right)\ .
\end{equation}
It is chosen to let the synchrotron tune be $\nu_s=\frac{\mu_s}{2\pi}$
for the case of zero dispersion at this cavity.  And the matrix of the
crab cavity is given by
\begin{equation}
\underline T_{\rm crab} = \left(\begin{array}{cccc}
1   & 0 & 0 & 0 \\
0   & 1 & \xi & 0 \\
0   & 0 & 1 & 0 \\
\xi & 0 & 0 & 1
\end{array}\right)\ .
\end{equation}
The one turn matrix just after the accelerating cavity is given by $\underline
T_{\rm cav}\underline T_{cc}$ when the crab cavity is switched off, and
the one turn matrix just before the crab cavity is given by
\begin{equation}
\underline T_{kk}^{{\rm tot}}=\underline T_{kc}\underline T_{cav}\underline
T_{cc}\underline T_{kc}^{-1}\underline T_{\rm crab}\ .
\end{equation}
All of these matrices are symplectic.

Synchro-beta resonances are driven by either dispersion at the
accelerating cavity, i.e. $\vec\eta_c\ne\vec 0$, or by a crab cavity
with $\xi\ne 0$. When $\vec\eta_c=\vec 0$ and $\xi=0$, we have a
decoupled case, and $\underline T_{kk}^{{\rm tot}}$ has the four
eigenvalues $e^{\pm i\mu_{x,s}}$. In general the eigenvalues are
determined by the characteristic equation
\begin{equation}
{\rm Det}(\underline T_{kk}^{{\rm tot}}-\lambda\cdot\underline 1) = 0\ ,
\label{eq:eigen}
\end{equation}
which gives
\begin{equation}
y^2+a_1 y + a_2 = 0\ ,\ \ 
y = \lambda+\frac{1}{\lambda}\ .
\end{equation}
Here $a_1$ is the coefficient for $\lambda$ in Eq.~(\ref{eq:eigen})
and $a_2+2$ is the coefficient for $\lambda^2$.

For stability, we need all four eigenvalues $\lambda$ to have unit
absolute values.  This requires that (A) $y=\lambda+\frac{1}{\lambda}$
is real so that $\lambda$ is on the unit circle whenever it is
complex. This condition requires
\begin{equation}
a_1^2-4a_2\ge 0\ .
\end{equation}
Furthermore we require (B) $|y|<2$ so that $\lambda$ is not real. This
requires
\begin{equation}
\pm a_1+\sqrt{a_1^2-4a_2}<4\ .
\end{equation}
This amounts to two conditions depending on the sign of $a_1$, to which
we refer as (B$_+$) and (B$_-$).

The first condition to (A) can easily be violated at first order
resonances $\mu\pm\mu_s=2\pi n$. Synchro-beta coupling can then move
two eigenvalues $\lambda$ toward each other. When two eigenvalues
become close, they can move away from the unit circle of $|\lambda|=1$
in the complex plane. Since the synchrotron phase advance $\mu_s$ is
close to 0, this condition becomes relevant when the horizontal tune
$\nu=\frac{\mu}{2\pi}$ is close to an integer.

The second condition (B) becomes relevant when the horizontal
tune is close to an integer or a half integer. The coupling can then
move $\lambda$ to real values.

In the following we compute conditions $(A)$ and $(B)$ for the
following cases:
\begin{itemize}
\item[A)] No crab cavity but with dispersion at the accelerating cavity,
i.e. $\xi=0$ and $\vec\eta_c\ne\vec 0$.
\item[B)] No dispersion at the crab cavity and at the accelerating cavity,
i.e. $\xi\ne 0$, $\vec\eta_k=\vec 0$ and $\vec\eta_c=\vec 0$.
\item[C)] Dispersion at the crab cavity but no dispersion at the
accelerating cavity, i.e. $\xi\ne 0$, $\vec\eta_k\ne\vec 0$ and
$\vec\eta_c=\vec 0$.
\end{itemize}
Case B is a special case of case C.

We will evaluate the stop-bands that occur close to integer and
half-integer values of the horizontal tune $\nu$ due to condition (A)
and (B) respectively. As an example we will use the parameters of
Tab.~\ref{tb:example} unless specified otherwise.  Typically the crab
cavity strength is given by the half crossing angle $\phi$ at the
interaction point via
\begin{equation}
\xi=\frac{2\phi}{\sqrt{\beta_k\beta_x^*}}\ ,
\end{equation}
where $\beta_x^*$ is the horizontal $\beta$ function at the
interaction point.  The range of the crab cavity strength used in the
following examples is $\xi\in[0,0.03]\frac{1}{{\rm m}}$. This range is
motivated by a crossing angle of about $\phi=30$mrad and beta
functions of about $\beta_k=50$m and $\beta_x^*=0.02$m. The range of
synchrotron tunes used in the following examples is
$\mu_s\in[-0.1,0.1]$.

\begin{table}
\begin{tabular}{lllllllllllll}
$\eta_k$     &=& 10m        &,\ $\eta_c$  &=& 10m&,\ 
$\eta'_k$    &=& 0          &,\ $\eta'_c$ &=&   0&,\\
$\beta_k$    &=& 10m        &,\ $\alpha_k$&=&   0&,\ 
$\beta_c$    &=& 10m        &,\ $\alpha_c$&=&   0&,\\
$\alpha$     &=&0.01        &,\ $R$       &=&100m&,\
$T_{56}^{kc}$&=& 0          &.            & &    &
\end{tabular}
\caption{Parameters for the stop-band examples. The numbers for the
dispersion apply in all examples where they are not specified
otherwise.}\label{tb:example}
\end{table}

\subsection{No crab cavity}

For $\xi=0 $ and $\vec\eta_c\ne\vec 0$, the coefficients $a_1$ and $a_2$
come out to be
\begin{eqnarray}
a_1^A &=& -2(\cos\mu+\cos\mu_s+2h_c\sin\mu\sin^2\frac{\mu_s}{2})\ ,
\label{eq:a1noxi}\\
a_2^A &=&  4(\cos\mu \cos\mu_s+2h_c\sin\mu\sin^2\frac{\mu_s}{2})\ .
\label{eq:a2noxi}
\end{eqnarray}
Since there is no crab cavity, the eigenvalues do not depend on the Twiss
parameters or phase advances with index $k$.

\subsection{No dispersion at both cavities}

The transport matrix for one turn that starts just before the crab
cavity depends on the phase advance between the accelerating and the
crab cavity and also on the time of flight term $T^{kc}_{56}$ between
these locations.  We express this term relative to the momentum
compaction as $r_{56}=\frac{T^{kc}_{56}}{2\pi R\alpha}$.  For $\xi\ne0 $
and $\vec\eta_k = \vec\eta_c =\vec 0$, the coefficients $a_1$ and $a_2$
come out to be
\begin{eqnarray}
a_1^B &=& -2(\cos\mu+\cos\mu_s)\ ,
\label{eq:a1nod}\\
a_2^B &=&  4\{\cos\mu \cos\mu_s\label{eq:a2nod}\\
    &+& \xi^2\beta_k\sin\mu\frac{\pi R\alpha}{2}
[1+4r_{56}(1+r_{56})\sin^2\frac{\mu_s}{2}]\}\ .
\nonumber
\end{eqnarray}
It is interesting to note that the eigenvalues do not depend on the
phase advance between the cavities.

\subsection{Dispersion at a crab cavity}

When there is a crab cavity, we want to analyze the leading order
effect, which is second order in the perturbations $\eta_k$,
$\eta'_k$, and $\xi$.  For $\xi\ne 0 $, $\vec\eta_k\ne\vec 0$, $a_1$
and $a_2$ come out to be
\begin{eqnarray}
a_1^C
&=&
a_1^B + 4\xi\eta_k(1+2r_{56})\sin^2\frac{\mu_s}{2} + 2\xi g_k\sin\mu\ ,
\label{eq:a1d}\\ 
a_2^C &=& a_2^B - 8\xi[(1+r_{56})\eta_k\cos\mu \sin^2\frac{\mu_s}{2}
+\frac{g_k}{2}\sin\mu\cos\mu_s]\nonumber\\
&+&
2 \xi^2 \{\eta_k^2(\cos\mu\cos\mu_s-1)\nonumber\\
&+&
2\eta_k[g_k(1+2r_{56})+h_k\eta_k)\sin\mu\sin^2\frac{\mu_s}{2}]\}\ .
\label{eq:a2d}
\end{eqnarray}
Here the short form $g_k=\alpha_k\eta_k+\beta_k\eta'_k$ and
$h_k=\frac{1}{2\pi
R\alpha}(\gamma_k\eta_k^2+2\alpha_k\eta_k\eta_k'+\beta_k\eta_k^{'2})$
were used.

\section{Horizontal Tune Close to Integers}
 
Condition (A) is relevant when the horizontal phase advance is close
to a given synchrotron phase advance $\mu_s$, which is usually much
smaller than $2\pi$, this equation leads to a stop-band for the
horizontal tune $\nu=\frac{\mu}{2\pi}$ close to every integer. The
boundary of stability is found by solving
\begin{equation}
a_1^2-4a_2=0
\end{equation}
for $\mu_s$.

One additionally has to check whether condition (B) is satisfied in
the regions which are declared as stable. For the parameters chosen
here, this is the case.

\subsection{No crab cavity}

For $\xi=0$, Eqs.~(\ref{eq:a1noxi}) and (\ref{eq:a2noxi}) for $a_1^A$
and $a_2^A$ lead to
\begin{eqnarray}
\cos\mu_s
&=&
\frac{1}{(1-h_c\sin\mu)^2}
(
\pm 4\sin^2\frac{\mu}{2}\sqrt{h_c\sin\mu}\\
&+&
\cos\mu-h_c(3-\cos\mu-h_c\sin\mu)\sin\mu
)
\ .
\label{eq:noxiex}
\end{eqnarray}
At the stop-band $\mu$ is close to $\mu_s$, which is usually small,
and we therefore expand to second order in the fractional part
$\delta\nu\in[-\frac{1}{2},\frac{1}{2}]$ of the horizontal tune
$\nu=\frac{\mu}{2\pi}$ leading to the approximate location of the
stop-band at
\begin{equation}
\mu_s=\pm(\delta\mu+h_c\sin^2\mu)
\end{equation}
and to the width of the stop-band
\begin{equation}
\Delta\mu_s=\pm\sqrt{h_c}\sin^{\frac{3}{2}}\mu\ .
\label{eq:noxiap0}
\end{equation}
Note that this only leads to a real stop-band width when $\nu$ is
slightly above an integer. Here and in all subsequent statements we
assume that the energy is above transition, i.e.~$\alpha>0$. Note that
the synchrotron tune above transition is negative. However, all
subsequent formulas depend on $\cos\mu_s$ only so that the sign of the
synchrotron tune does not matter.

\begin{figure}[ht!]
\begin{center}
\begin{minipage}{0.8\linewidth}
\includegraphics[width=\linewidth,clip]{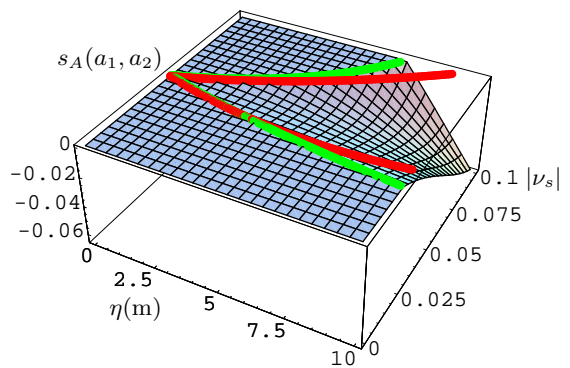}
\setlength{\unitlength}{\linewidth}
  \begin{picture}(1,0)(0,0)
    \put(0.10,0.70) {$s_A(a_1,a_2)$}
    \put(0.20,0.22) {$\eta$(m)}
    \put(1.00,0.47) {$|\nu_s|$}
  \end{picture}
\end{minipage}
\end{center}
  \caption{Unstable region caused by dispersion at a cavity for $\nu$
  close to an integer. Light green: border of stability. Dark red:
  border of stability from the approximate Eq.~\ref{eq:noxiap0}.}
  \label{fg:d0}
\end{figure}

Figure \ref{fg:d0} shows the unstable region caused by dispersion in a
cavity. The parameters of Tab.~\ref{tb:example} were used. The valley
of instability shows $a_1^2-4a_2$ in the region where it is negative,
i.e.~ the stability function $s_A(a_1,a_2)={\rm Min}(a_1^2-4a_2,0)$ is
plotted.  The light green line indicates the border of stability
computed by Eq.~(\ref{eq:noxiex}). The approximation in
Eq.~(\ref{eq:noxiap0}) leads to the dark red curve. It is apparent
that the approximation is very good for $\eta<5$m.

\subsection{No dispersion at both cavities}
For $\xi\ne 0$, $\vec\eta_c=\vec\eta_k=\vec 0$ we obtain
from Eqs.~(\ref{eq:a1nod}) and (\ref{eq:a2nod}) for $a_1^B$ and $a_2^B$
an equation for the boundary of instability for $\cos\mu_s$.
For the example of $r_{56}=0$ it is
\begin{equation}
\cos\mu_s=\pm\cos\mu\pm\xi\sqrt{2\pi R\alpha_k\beta_k\sin\mu}\ .
\end{equation}

To leading order in $\xi$ one obtains the approximate location and
width of the stop-band,
\begin{eqnarray}
\mu_s &=& \pm\delta\mu\ ,\\
\Delta\mu_s &=& \pm\xi
\sqrt{\frac{2\pi R\alpha\beta_k}{\sin\mu}}
\sqrt{1 + 4r_{56}(1+r_{56})\sin^2\frac{\mu}{2}}\ .
\label{eq:nodap0}
\end{eqnarray}

\begin{figure}[ht!]
  \begin{center}
\begin{minipage}{0.8\linewidth}
\includegraphics[width=\linewidth,clip]{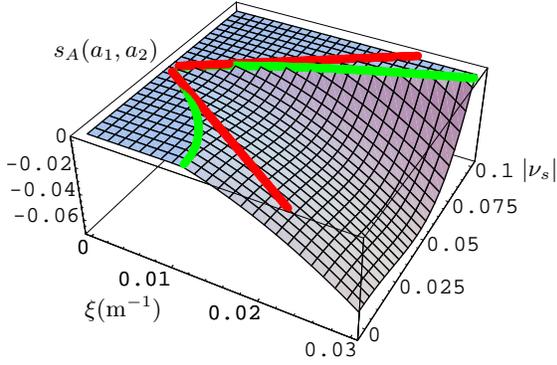}\\
\setlength{\unitlength}{\linewidth}
  \begin{picture}(1,0)(0,0)
    \put(0.1,0.70) {$s_A(a_1,a_2)$}
    \put(.16,.2) {$\xi$(m$^{-1}$)}
    \put(1.00,.47) {$|\nu_s|$}
  \end{picture}
\end{minipage}
\end{center}
  \caption{Unstable region caused by a crab cavity. Light green:
  border of stability, Dark Red: approximation to this border.}
  \label{fg:xi0}
\end{figure}

Figure \ref{fg:xi0} shows the unstable region. Again the light green
curves show the border of stability and the dark red curves indicate
the approximation of Eq.(\ref{eq:nodap0}), which apparently is very
good for $\xi<0.007\frac{1}{{\rm m}}$.

\subsection{Dispersion at a crab cavity}

If the dispersion in the crab cavity is not matched to a small value,
it will typically be as large as several meters. Since this is not a
small perturbation, we only linearize in $\xi$ and in $\delta\mu$.

When using Eqs.~(\ref{eq:a1d}) and (\ref{eq:a2d}) for solving
$a_1^{C2}-4a_2^C = 0$ for $\mu_s$ and subsequently linearizes in $\xi$
and $\delta\mu$ one obtains
\begin{equation}
\mu_s
=
\pm \{\delta\mu + \xi[g_k-\eta_k(1+2r_{56})\frac{\sin\mu}{2}]\}\ .
\end{equation}

For the width of the stop-band one obtains
\begin{equation}
\Delta\mu_s=\pm \xi\sqrt{\frac{2\pi R\alpha\beta_k}{\sin\mu}}
( 1-\frac{\eta_k^2}{2\pi R\alpha\beta_k}\sin\mu )\ .
\label{eq:xid0}
\end{equation}

It is interesting that no real stop-band width exists when $\sin\mu$
is negative, and the stop-band only occurs above integer horizontal
tunes. Higher order terms of the stop-band width expansion depend on
the time of flight term $r_{56}$, $\alpha_k$ and $\eta'_k$.

Now we can evaluate whether dispersion at the crab cavity has a
negative influence on the stop-band width.  Since there is only a
stop-band for $\sin\mu>0$, it turns out that in
Eq.~(\ref{eq:xid0}) a dispersion at the crab cavity reduces
the stop-band width, as long as $\eta_k^2$ is smaller than $\frac{2\pi
R\alpha\beta_k}{\sin\mu}$.

\begin{figure}[ht!]
\begin{center}
\begin{minipage}{0.8\linewidth}
\includegraphics[width=\linewidth,clip]{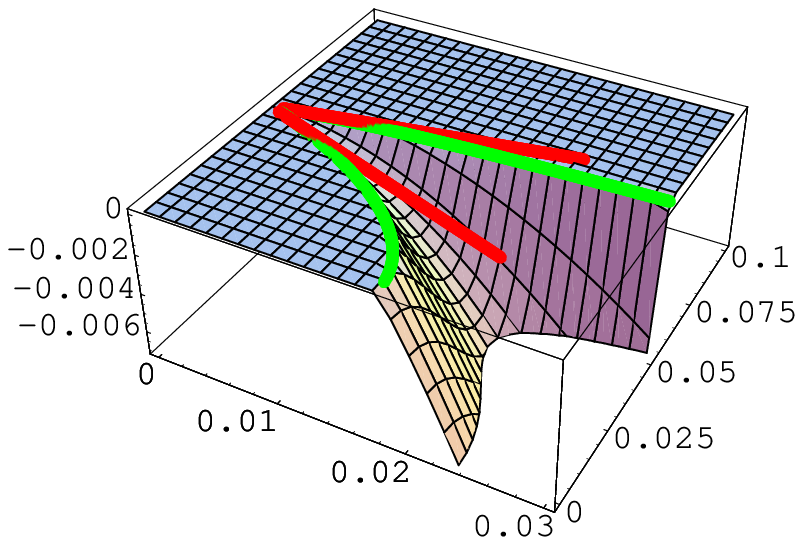}
\setlength{\unitlength}{\linewidth}
  \begin{picture}(1,0)(0,0)
    \put(0.1,0.70) {$s_A(a_1,a_2)$}
    \put(.16,.2) {$\xi$(m$^{-1}$)}
    \put(1.00,.47) {$|\nu_s|$}
  \end{picture}
\end{minipage}
\begin{minipage}{0.8\linewidth}
\includegraphics[width=\linewidth,clip]{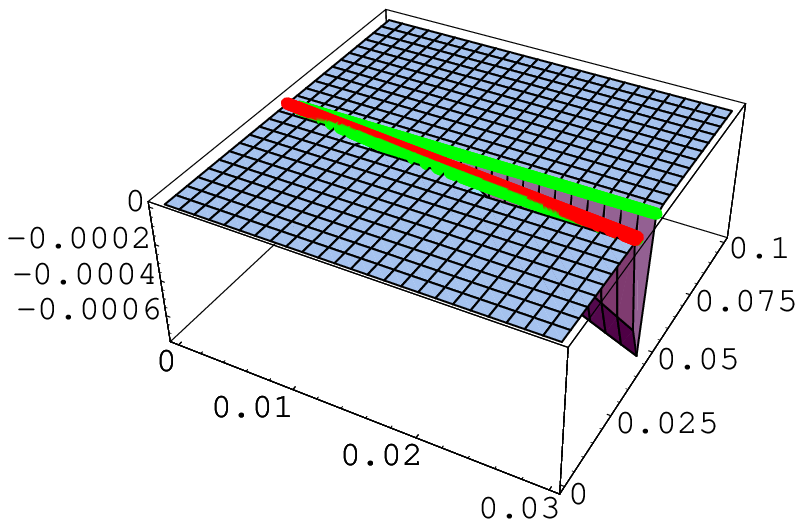}
\setlength{\unitlength}{\linewidth}
  \begin{picture}(1,0)(0,0)
    \put(0.1,0.70) {$s_A(a_1,a_2)$}
    \put(.16,.2) {$\xi$(m$^{-1}$)}
    \put(1.00,.47) {$|\nu_s|$}
  \end{picture}
\end{minipage}
\end{center}
  \caption{Unstable region caused by a crab cavity for a horizontal
  slightly above an integer. Top: $\eta_k=11$m, Bottom:
  $\eta_k=14.32$m.  Light green: border of stability. Dark red: border
  of stability from the approximate Eq.~\ref{eq:xid0}.}
  \label{fg:xid0}
\end{figure}

Figure \ref{fg:xid0}~(bottom) shows the unstable region bordered by a
light green curve for the optimal value of $\eta_k=\sqrt{\frac{2\pi
R\alpha\beta_k}{\sin\mu}}=14.32$m. The approximate Eq.~(\ref{eq:xid0})
indicated by the dark red curve evaluates to zero stop-band width for
this dispersion. In fact, the stop-band is extremely narrow but due to
higher order terms it does not have zero width.  But
Fig.~\ref{fg:xid0}~(top) for $\eta_k=11$m shows that the approximation
works quite reliably for non-optimal dispersions.

The other parameters were chosen as specified in
Tab.~\ref{tb:example}. It is evident that the dispersion reduces the
width of the stop-band greatly, and it could be recommended to use the
dispersion at the crab cavity to reduce the stop-band width. However,
the B-factories at SLAC and KEK as well as other $e^+/e^-$ colliders
like CESR have horizontal tunes which are close to a half integer and
therefore the subsequent section about stop-bands at half-integer
tunes is more relevant for these applications.

\section{Horizontal Tune Close to Half Integers}
 
When using condition (A) for determining stability, one additionally
has to check whether condition (B) is satisfied in the regions which
are declared stable by condition (A). For the parameters of the
examples shown above, this is the case. When, as in the case of the
B-factories or other $e^+/e^-$ colliders the horizontal tune is close
to a half integer, whereas the synchrotron tune is small, the coupling
strength will not bring two eigenvalues $\lambda$ together.  It can
however happen that one of the eigenvalues becomes real, and it is
here that condition (B) becomes important. In the following
examples we therefore use $\nu=8.49$ or $\nu=8.51$.

The borders of stability are found by solving
\begin{equation}
\pm a_1+\sqrt{a_1^2+2a_2}=4
\label{eq:borderB}
\end{equation}
for $\mu_s$.

\subsection{No crab cavity}
Using the coefficients $a_1^A$ and $a_2^A$ from Eqs.~(\ref{eq:a1noxi})
and (\ref{eq:a2noxi}), Eq.~\ref{eq:borderB} yields the border of
stability, and for $a_1^A>0$ one obtains
\begin{equation}
\cos\mu_s = -\frac{\cos\frac{\mu}{2} +
  2h_c\sin\frac{\mu}{2}}{\cos\frac{\mu}{2} - 2h_c\sin\frac{\mu}{2}}\ .
\end{equation}
For small dispersion at the cavity one obtains
$\cos\mu_s\approx-1-4\tan\frac{\mu}{2}$, which cannot be satisfied for
small synchrotron tunes. For $a_1^A<0$ one obtains only $\cos\mu_s=1$
so that there is no unstable region close to half-integer tunes due to
dispersion at an accelerating cavity.

\subsection{No dispersion at both cavities}

From $a_1^B$ and $a_2^B$ in Eqs.~(\ref{eq:a1nod}) and (\ref{eq:a2nod})
we obtain
\begin{equation}
\cos\mu_s=1+\xi^2\frac{1}{2}\pi R\alpha\beta_k\cot\frac{\mu}{2}\ .
\end{equation}
To first order in $\xi$ this leads to a resonance at $\mu_s=0$ of width
\begin{equation}
\Delta\mu_s = \pm \xi\sqrt{-\pi R\alpha \beta_k \cot\frac{\mu}{2}}\ .
\label{eq:nodap1}
\end{equation}
The stop-band in $\mu_s$ thus only appears when the horizontal tune is
slightly above the half integer. An expansion to higher orders depends
on $r_{56}$, $\alpha_n$ and $\eta'_k$.

\begin{figure}[ht!]
  \begin{center}
\begin{minipage}{0.8\linewidth}
\includegraphics[width=\linewidth,clip]{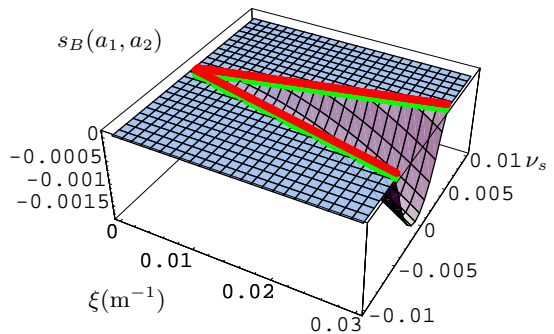}
  \setlength{\unitlength}{\linewidth}
  \begin{picture}(1,0)(0,0)
    \put(0.1,0.70) {$s_B(a_1,a_2)$}
    \put(0.16,0.2) {$\xi$(m$^{-1}$)}
    \put(1.00,0.47) {$\nu_s$}
  \end{picture}
\end{minipage}
\end{center}
  \caption{Unstable region caused by a crab cavity for small
  distances $\delta\nu$ above a half-integer tune. Light green:
  border of stability.}
  \label{fg:xi1}
\end{figure}

In Fig.~\ref{fg:xi1} the valley of instability is shown by plotting
$s_B(a_1,a_2)={\rm Min}(4-|a_1|-\sqrt{a_1^2-4a_2},0)$
for $\nu=8.51$. The approximation in Eq.~(\ref{eq:nodap1}) is shown
together with the stop-band region. The agreement is very good.  Note
that the range of $\nu_s$ is reduced by a factor of 10 from compared
to other graphs. Here the forbidden region is up to 0.01 in tune
space. This is currently not critical for the B-factories but could in
the future restrict the possibilities of moving the horizontal tune
even closer to the half integer in B-factories.

\subsection{Dispersion at a crab cavity}

Using $a_1^C$ and $a_2^C$ in Eqs.~(\ref{eq:a1d}) and (\ref{eq:a2d}) to
find the border of stability for $\mu_s$, and then linearizing in
$\xi$ leads to the stop-band width
\begin{equation}
\Delta\mu_s = \pm \xi\sqrt{\eta_k^2-\pi
R\alpha\beta_k\cot\frac{\mu}{2}}
\label{eq:xid1}
\end{equation}
and the stop-bands center at $\mu_s=0$.  This result comes from the
condition (B$_-$), $-a_1^C+ \sqrt{a_1^{2C}-4a_2^{C2}} = 4$, since the
condition (B$_+$) requires $\nu_s=0.5$ for $\xi=0$.  Higher order
expansions again depend on $r_{56}$, $\alpha_k$ and $\eta'_k$.

Equation (\ref{eq:xid1}) indicates that for tunes where there is a
stop-band in the case of $\eta_k=0$, i.e. for $\nu$ above a half
integer where $\cot\frac{\mu}{2}<0$, the stop-band becomes wider by
introducing dispersion at the crab cavity. Furthermore, there is now
also a stop-band at tunes below a half integer where
$\cot\frac{\mu}{2}>0$ when the dispersion is large enough.

\begin{figure}[ht!]
\begin{center}
\begin{minipage}{0.49\linewidth}
\includegraphics[width=\linewidth,clip]{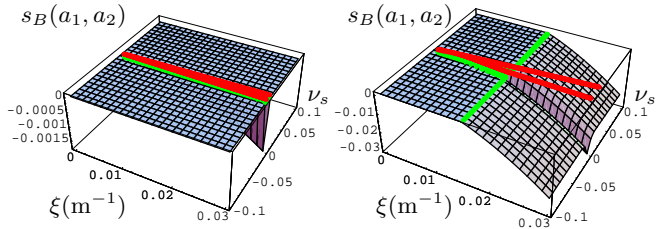}
\setlength{\unitlength}{\linewidth}
  \begin{picture}(1,0)(0,0)
    \put(0.05,0.80) {$s_B(a_1,a_2)$}
    \put(.14,.2) {$\xi$(m$^{-1}$)}
    \put(0.95,.55) {$\nu_s$}
  \end{picture}
\end{minipage}
\begin{minipage}{0.49\linewidth}
\includegraphics[width=\linewidth,clip]{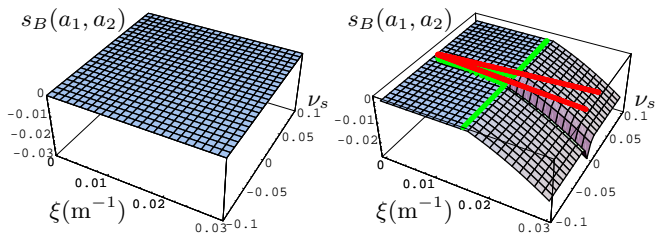}
\setlength{\unitlength}{\linewidth}
  \begin{picture}(1,0)(0,0)
    \put(0.05,0.80) {$s_B(a_1,a_2)$}
    \put(.14,.2) {$\xi$(m$^{-1}$)}
    \put(0.95,.55) {$\nu_s$}
  \end{picture}
\end{minipage}
\end{center}
  \caption{Unstable region caused by a crab cavity for $\nu$ close to a
  half integer with $\cot\frac{\mu}{2}<0$. Left: $\eta_k=0$, Right:
  $\eta_k=4$m.  Light green: border of stability. Dark red: border of
  stability from the approximate Eq.~\ref{eq:xid1}.}
  \label{fg:xid1}
\end{figure}

\begin{figure}[ht!]
\begin{center}
\begin{minipage}{0.49\linewidth}
\includegraphics[width=\linewidth,clip]{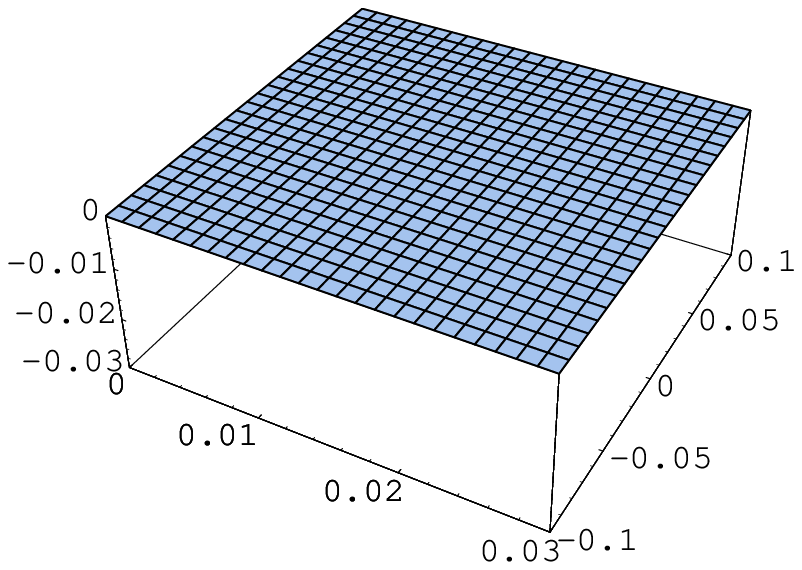}
\setlength{\unitlength}{\linewidth}
  \begin{picture}(1,0)(0,0)
    \put(0.05,0.80) {$s_B(a_1,a_2)$}
    \put(.14,.2) {$\xi$(m$^{-1}$)}
    \put(.95,.55) {$\nu_s$}
  \end{picture}
\end{minipage}
\begin{minipage}{0.49\linewidth}
\includegraphics[width=\linewidth,clip]{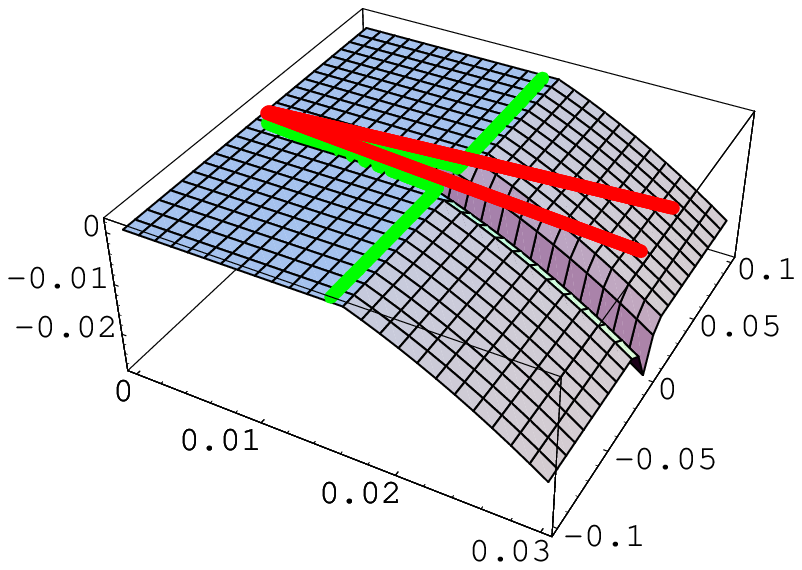}
\setlength{\unitlength}{\linewidth}
  \begin{picture}(1,0)(0,0)
    \put(0.05,0.80) {$s_B(a_1,a_2)$}
    \put(.14,.2) {$\xi$(m$^{-1}$)}
    \put(.95,.55) {$\nu_s$}
  \end{picture}
\end{minipage}
\end{center}
  \caption{Unstable region caused by a crab cavity for $\nu$ close to a
  half integer with $\cot\frac{\mu}{2}>0$. Left: $\eta_k=0$, Right:
  $\eta_k=4$m.  Light green: border of stability. Dark red: border of
  stability from the approximate Eq.~\ref{eq:xid1}.}
  \label{fg:xid1m}
\end{figure}

Figure~\ref{fg:xid1} shows the unstable region for $\mu_s$ with
$\eta_k=0$ (left) and $\eta_k=4$m (right) for a tune of $\nu=8.51$
where $\cot\frac{\mu}{2}<0$. Especially for larger values of $\xi$ the
difference is very clear.  It is even clearer when investigating the
unstable region for a tune of $\nu=8.49$ in Fig.~\ref{fg:xid1m}. For
$\eta_k=0$ (left) there is no stop-band while for $\eta_k=4$m (right)
the stop-band is substantial.

 The border of stability has two sections, one is well approximated by
Eq.~(\ref{eq:xid1}) and one is not. The former one comes from
condition (B$_-$) on which Eq.~(\ref{eq:xid1}) is based and the latter
one comes from condition (B$_+$).

\begin{figure}[ht!]
\begin{center}
\begin{minipage}{0.8\linewidth}
\includegraphics[width=\linewidth,clip]{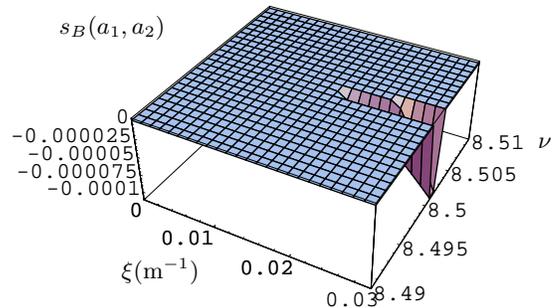}
\setlength{\unitlength}{\linewidth}
  \begin{picture}(1,0)(0,0)
    \put(0.1,0.70) {$s_B(a_1,a_2)$}
    \put(.22,.23) {$\xi$(m$^{-1}$)}
    \put(1.02,.48) {$\nu$}
  \end{picture}
\end{minipage}
\begin{minipage}{0.8\linewidth}
\includegraphics[width=\linewidth,clip]{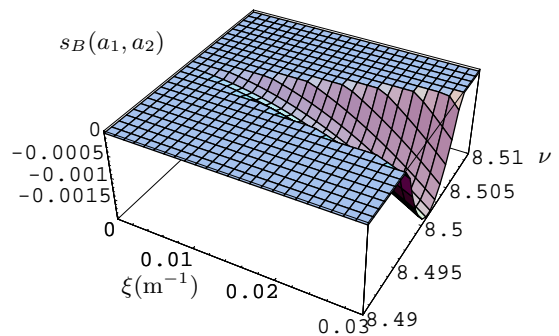}
\setlength{\unitlength}{\linewidth}
  \begin{picture}(1,0)(0,0)
    \put(0.1,0.70) {$s_B(a_1,a_2)$}
    \put(.22,.23) {$\xi$(m$^{-1}$)}
    \put(1.02,.48) {$\nu$}
  \end{picture}
\end{minipage}
\end{center}
  \caption{Unstable region caused by a crab cavity for
  $\nu_s=0.1$. Top: $\eta_k=0$, Bottom: $\eta_k=1$m.  Light green:
  border of stability. Dark red: border of stability from the
  approximate Eq.~\ref{eq:xid1}.}
  \label{fg:xid1x}
\end{figure}

The strength of this effect is nicely seen for a dispersion of only
$\eta_k=1$m when plotting the stop-band of the horizontal tune $\nu$
for a fixed synchrotron tune of $\nu_s=\frac{\mu_s}{2\pi}=0.1$ in
Fig.~\ref{fg:xid1x}. For $\eta_k=0$ (top) the stop-band is located
only on one side of the half integer. For $\eta_k=1$m (bottom) the $\nu_s$
stop-band is not only much wider, it also extends to both sides of the
half integer.

\section{A pair of crab cavities}

While we here want to restrict ourselves to the proposed insertion of
one crab cavity in a ring, it should be pointed out that a pair of
crab cavities can strongly reduce the width of the stop-bands when the
two crab cavities have betatron phases which differ by odd multiples
of $\pi$.  While this would cancel the first order synchro-beta
coupling of the crab cavity exactly when the time of flight term
$m_{56}$ between them vanishes, the beam-beam focusing at the
interaction point will spoil this exact cancellation to some
extent.

It can be shown that, when a beam-beam focusing is included at
the IP, the effect can be represented as a single kick at the entrance
crab cavity by a matrix
\begin{equation}
T_{\rm crab} = \underline 1 + 4\pi\Delta\nu \beta_k\left( \begin{array}{cccc}
0 & 1 & \xi & 0 \\
0 & 0 & 0 & 0 \\
0 & 0 & 0 & 0 \\
0 & -\xi & -\xi^2 & 0
\end{array}\right)
\end{equation}
where $\Delta \nu$ is the beam-beam tune shift parameter, and
$m_{56}=0$ was assumed for the transport matrix between the cavities.
Furthermore we assumed $\vec\eta_k=\vec 0$ and $\alpha_k=0$.  As it should,
when $\Delta \nu =0$, the two crab cavities have perfect cancellation
and this matrix becomes a unit matrix.

\begin{figure}[ht!]
\begin{center}
\begin{minipage}{0.8\linewidth}
\includegraphics*[width=\linewidth]{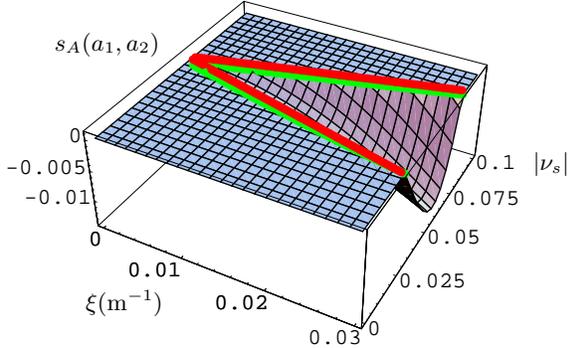}
\setlength{\unitlength}{\linewidth}
  \begin{picture}(1,0)(0,0)
    \put(0.1,0.70) {$s_A(a_1,a_2)$}
    \put(.16,.2) {$\xi$(m$^{-1}$)}
    \put(1.02,.47) {$|\nu_s|$}
  \end{picture}
\end{minipage}
\end{center}
  \caption{Unstable region caused by two crab cavities, $\pi$ apart in
  betatron phase for $\nu=8.025$.  Light green: border of
  stability. Dark red: border of stability from the approximate
  Eq.~\ref{eq:xixi0}.}
  \label{fg:xixi0}
\end{figure}

The analysis can now be repeated with this new matrix for the crab
cavity.  It is found that the stop-band is relatively narrow. For the
above numerical example with $\xi\ne 0$, $\vec\eta_c=\vec\eta_k=\vec 0$ and
$\nu = 8.025$ the unstable region is shown in Fig.~\ref{fg:xixi0} for
$\Delta\nu=0.05$.  Note that the stop-band is narrower and that the
unstable valley is shallower than in the corresponding
Fig.~\ref{fg:xi0}.

The width of the stop-band at $\mu\pm\mu_s=n2\pi$ with a pair of crab
cavities, for the case when there is no dispersion at the crab
cavities and at the accelerating cavity, is given by
\begin{eqnarray}
\mu_s &=& \pm(\delta\mu+2\pi\Delta\nu)\ ,\\
\Delta\mu_s
&=& \pm \xi\Delta\nu 4\pi
\label{eq:xixi0}\\
&\times& \sqrt{\frac{2\pi R\alpha\beta_k}{\sin\mu}}
\sqrt{1+4r_{56}(1+r_{56}\sin^2\frac{\mu}{2})}\nonumber
\end{eqnarray}
which is a factor $4\pi\Delta\nu$ smaller than that due to a single
crab cavity in Eq.~(\ref{eq:nodap0}). Note however that $\Delta\nu$
may become larger than $0.05$ in high luminosity operations.  Again,
the resonance only appears above the integer.

\begin{figure}[t!]
\begin{center}
\begin{minipage}{0.8\linewidth}
\includegraphics*[width=\linewidth]{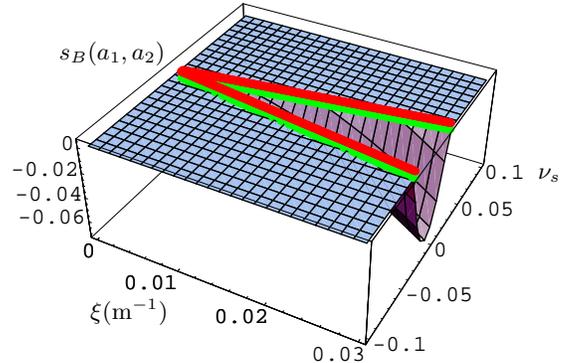}
\setlength{\unitlength}{\linewidth}
  \begin{picture}(1,0)(0,0)
    \put(0.1,0.70) {$s_B(a_1,a_2)$}
    \put(.16,.2) {$\xi$(m$^{-1}$)}
    \put(1.02,.47) {$\nu_s$}
  \end{picture}
\end{minipage}
\end{center}
  \caption{Unstable region caused by two crab cavities, $\pi$ apart in
  betatron phase for $\nu=8.51$.  Light green: border of
  stability. Dark red: border of stability from the approximate
  Eq.~\ref{eq:xixi1}.}
  \label{fg:xixi1}
\end{figure}

When the tune is close to a half integer, the border of stability for
$\mu_s$ becomes
\begin{equation}
\Delta\mu_s = \pm\xi\sqrt{4\pi\Delta\nu}\sqrt{2\pi R\alpha\beta_k}\ .
\label{eq:xixi1}
\end{equation}
This differs from the corresponding Eq.(\ref{eq:nodap1}) approximately
by the factor $2\sqrt{\frac{4\pi\Delta\nu}{\delta\mu}}$. An example
for $\nu=8.51$ is given in Fig.~\ref{fg:xixi1}. For the example of
$\Delta\nu=0.05$ and $\delta\nu=0.01$, the valley of instability is
wider and deeper than in the corresponding Fig.~\ref{fg:xi1}.

\section{Conclusion}
The linear synchro-beta coupling effects have been analyzed for a
storage ring with a crab cavity or a crab cavity pair.  When one crab
cavity is to be installed in a storage ring, the dispersion at the
crab cavity can be non-zero or can even be used to reduce synchro-beta
stop-bands. When the tune is close to a half integer, the dispersion
should be matched to a small value at the crab cavity however.  In
both cases it should be noted that the stop-band due to synchro-beta
coupling is larger than that due to a large dispersion at an
accelerating cavity. This effect can be reduced when two crab cavities
are used. The advantage is limited, however, when the ring is operated
with a large beam-beam tune shift.


\begin{thebibliography}{9}

\bibitem{piwinski77} A.~Piwinski, IEEE Trans.~on Nucl.~Sci.~NS-24, 1480 (1977)

\bibitem{piwinski85} A.~Piwinski, IEEE Trans.~on Nucl.~Sci.~NS-32, 2240 (1985)

\bibitem{palmer88} R.~Palmer, SLAC-PUB-4707 (1988)

\bibitem{oide89} K. Oide and K. Yokoya, Phys. Rev. A40, 315 (1989)

\bibitem{Padamsee:1991wv}
H.~Padamsee, P.~Barnes, C.~Chen, J.~Kirchgessner, D.~Moffat, D.~Rubin,
Y.~Samed, J.~Sears, Q.S.~Shu, M.~Tigner, D.~Zu,
CLNS-91-1075, Cornell and
Proceedings PAC91, San Francisco/CA (1991)

\bibitem{Akai:2004np} K.~Akai and Y.~Morita,
Report KEK-PREPRINT-2003-123 (2003)

\bibitem{krishnagopal89}
S. Krishnagopal and R. Siemann, CLNS 89/967, Cornell (1989)

\bibitem{handbook} Handboodk of Accelerator Physics and Engeneering,
p.~51, Editors A.~W.~Chao, M.~Tigner, Word Scientific Publishing
Co.~Pte.~Ltd.~(2002)

\end{thebibliography}
\end{document}